\documentclass[aps,prb,reprint,amsmath,amssymb]{revtex4-2}
\usepackage{siunitx, tikz}
\usepackage{amsmath, amsthm, amssymb, graphicx, subfigure, braket, mhchem}

\usepackage[bookmarks=true, colorlinks, citecolor=blue, linkcolor=blue, urlcolor=blue]{hyperref}
\usepackage{bm}
\usepackage{color}

\begin{document}
\preprint{APS/123-QED}

\title{The generation and detection of the spin-valley-polarization in semi-Dirac materials
}


\author{Yupeng Huang$^{1}$}
\author{R. Shen$^{1,2}$}
\email{shen@nju.edu.cn}
\affiliation{$^{1}$National Laboratory of Solid State Microstructures and School of Physics, Nanjing University, Nanjing, 210093, China\\
$^{2}$ Collaborative Innovation Center of Advanced Microstructures, Nanjing University, Nanjing, 210093, China}


\date{\today}

\begin{abstract}
We investigated the transport properties in a normal metal/ferromagnet/normal metal/superconductor junction based on semi-Dirac materials with inverted energy gap. With a scattering matrix approach, we show that the electron transport in the junction is spin-valley-polarized due to the ferromagnetic exchange energy. It is also shown that the Andreev reflection is strongly suppressed, which is a clear experimental signal for the spin-valley-polarization in semi-Dirac materials.
\end{abstract}


\maketitle

\section{Introduction}
Valleytronics \cite{PhysRevLett.99.236809, rycerz2007valley, schaibley2016valleytronics, PhysRevB.77.205416, PhysRevLett.97.186404} is a concept that the electrons can be used to encode and process information by manipulating the valley degree of freedom. Successful fabrication of two-dimensional materials, such as graphene \cite{doi:10.1126/science.1102896} and transition metal dichalcogenides \cite{liu2019valleytronics, PhysRevB.90.195428, rivera2018interlayer, ye2016electrical, bawden2016spin, PhysRevLett.117.187401}, has facilitated the development of the valleytronics, leading to emerging applications, such as the valley Hall effect \cite{PhysRevLett.99.236809, PhysRevB.77.235406, doi:10.1126/science.1254966, shimazaki2015generation, doi:10.1126/science.1250140, PhysRevB.106.094503}, the valley Zeeman effect \cite{srivastava2015valley, PhysRevB.91.075433, lyons2019valley, PhysRevResearch.2.033256}, and the valley polarization \cite{zeng2012valley, mak2012control}. The valley polarization in graphene can be detected by the valley Hall effect \cite{doi:10.1126/science.1254966, shimazaki2015generation} and the Andreev reflection (AR) \cite{PhysRevB.96.245414}. Several approaches to generate the valley polarization in graphene have been proposed, such as introducing a line defect \cite{PhysRevLett.106.136806}, a ballistic point contact with zigzag edges \cite{rycerz2007valley}, and electric gate pontential in bilayer graphene \cite{PhysRevApplied.11.044033}. The valley polarization in semi-Dirac materials  can be generated by applying the gate voltage \cite{PhysRevB.96.245410}.

 The semi-Dirac material with inverted energy gap is characterized by anisotropic energy dispersion, with the conduction band and the valence band meeting at two nonequivalent Dirac points near which the electronic states have chiral pseudospins and Berry's phase \cite{PhysRevB.80.153412, montambaux2009universal, doi:10.1021/acs.nanolett.5b04106}. In the low-energy regime, the energy spectrum is composed of two Dirac cones. When the energy is beyond the inverted gap, two valleys merge into a single Fermi surface. Such energy dispersion is a merging Dirac cones system, which can be produced experimentally in the honeycomb optical lattice \cite{PhysRevB.80.153412, PhysRevLett.98.260402}, the \ce{TiO2}/\ce{VO2} heterostructures \cite{PhysRevLett.103.016402}, and the phosphorus with situ deposition of \ce{K} or \ce{Rb} atoms \cite{doi:10.1126/science.aaa6486}. Owing to its peculiar properties, the semi-Dirac materials with inverted gap have been investigated in valleytronics for potential applications \cite{PhysRevB.96.245410, PhysRevB.96.045424, jung2020black, Rostamzadeh_2022}. The transport of the electron from one valley can be inhibited when its pseudospin mismatches that of a gate-controlled scattering region, leading to a pseudospin-assisted valley-contrasting transport \cite{PhysRevB.96.245410}.

Inspired by the connection between semi-Dirac materials and the valley polarization, in this work, we propose that the transport in a normal metal/ferromagnet/normal metal (NFN) structure based on semi-Dirac materials is spin-valley-polarized due to the valley-selective tunneling, i.e., the currents in two valleys have opposite spin polarizations. The spin polarizations in the two valleys can be switched by reversing the direction of magnetization in the ferromagnetic layer. We further demonstrate that the AR in a semi-Dirac-based normal metal/ferromagnet/normal metal/superconductor (NFNS) junction is strongly suppressed due to the inhibition of the transport of the reflected hole, hence the spin-valley polarization can be detected by measuring the differential conductance in the NFNS junction.

The paper is organized as follows. In section II, the model Hamiltonian of the NFNS junction based on semi-Dirac materials is given. The transport coefficients are calculated using a scattering matrix approach. The numerical results and discussions are given in section III. A brief conclusion is given in section IV.

\section{Method}
The Hamiltonian of the anisotropic semi-Dirac system with inverted energy gap can be described as \cite{PhysRevB.80.153412, montambaux2009universal}
\begin{equation}
H_0(\bm K)=\begin{pmatrix}
0 & \dfrac{\hbar^2K_x^2}{2m^*}-\tilde{\Delta}_g-i \hbar v K_y \\
\dfrac{\hbar^2K_x^2}{2m^*}-\tilde{\Delta}_g+i \hbar v K_y & 0
\end{pmatrix},\label{Eq:1}
\end{equation}
where $\bm K=(K_x, K_y)$ is the wave vector, $\tilde{\Delta}_g>0$ is the inverted energy gap, $m^*$ is the effective mass along $K_x$ direction, and $v$ is the Dirac velocity along $K_y$ direction. With the momentum scale $p_0\equiv2m^*v$, the energy scale $E_0\equiv p_0^2/(2m^*)$, and the length scale $l_0\equiv\hbar/p_0$ \cite{PhysRevB.86.075124}, one can define the dimensionless quantities as $\bm k=\hbar\bm K/p_0$, $\Delta_g=\tilde{\Delta}_g/E_0$, $\mathcal{H}_0(\bm k)=H_0(\bm K)/E_0$. By substituting $\bm k$ with $-i\nabla_{\bm r}$, we get the effective Hamiltonian $\mathcal{H}_0(\bm r)$ for the semi-Dirac system in real space as
\begin{equation}
\mathcal{H}_0(\bm r)=\begin{pmatrix}
0 & -\partial_x^2-\Delta_g-i k_y \\
-\partial_x^2-\Delta_g+i k_y & 0
\end{pmatrix}.\label{Eq:2}
\end{equation}

\begin{figure}[htb]

\begin{tikzpicture}
\draw[very thick,dashed](-0.2,-0.9) -- (5.5,-0.9);
\draw[thick](0,0) parabola bend (1,-1.5) (2,0);
\draw[thick](0,-1.8) parabola bend (1,-0.3) (2,-1.8);
\draw[domain=3:3.32,red,thick] plot(\x,{(\x-4.25)^2/0.8-2});
\draw[domain=5.18:5.5,red,thick] plot(\x,{(\x-4.25)^2/0.8-2});
\draw[domain=3:3.32,blue,dashed,thick] plot(\x,{-(\x-4.25)^2/0.8+0.2});
\draw[domain=5.18:5.5,blue,dashed,thick] plot(\x,{-(\x-4.25)^2/0.8+0.2});

\filldraw [black] (7,-0.7) circle (1.6pt);
\draw[very thick][->](6.2,-0.7)--(6.8,-0.7);
\filldraw[fill=white,draw=black] (7,-1.1) circle (1.8pt);
\draw[very thick][<-](6.2,-1.1)--(6.8,-1.1);

\draw[thick](7.6,0.4)--(7.6,-2.2);
\draw (7.6,-2.3) node {$x=L$};
\draw[thick](2.625,0.4)--(2.625,-2.2);
\draw (2.625,-2.3) node {$x=-L_f$};
\draw[thick](5.7,0.4)--(5.7,-2.2);
\draw (5.7,-2.3) node {$x=0$};
\draw (6.5,-1.11) node{$\times$};

\filldraw [black] (1.74,-0.7) circle (1.2pt);
\draw (2.02,-0.7) node {$a_K$};
\filldraw [black] (1.52,-0.7) circle (1.2pt);
\draw (1.25,-0.7) node {$b_K$};
\filldraw [black] (0.27,-0.7) circle (1.2pt);
\draw (0.02,-0.7) node {$b_{K^{\prime}}$};
\filldraw[fill=white,draw=black] (1.73,-1.1) circle (1.4pt);
\draw (2.02,-1.1) node {$b_K^h$};
\filldraw[fill=white,draw=black] (0.48,-1.1) circle (1.4pt);
\draw (0.82,-1.1) node {$b^h_{K^{\prime}}$};

\draw (1,0.2) node {N$_1$};
\draw (4.25,0.2) node {F};
\draw (6.5,0.2) node {N$_2$};
\draw (8.2,-0.9) node {S};
\end{tikzpicture}
\caption{\label{Fig:1}The schematic band structures of the NFNS junction. The dashed line denotes the Fermi level, the black dots and the white circles denote the electrons and the holes respectively. In the F region, the spin-up conduction band above the Fermi level is plotted by red solid lines, while the spin-down valence band below the Fermi level is plotted by blue dashed lines.
}
\end{figure}
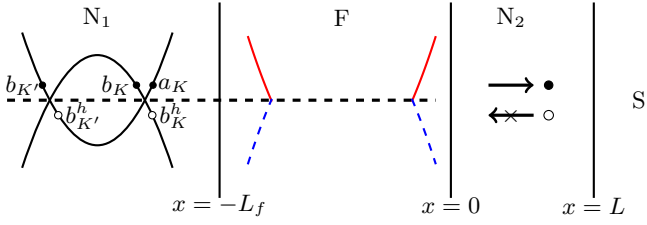

We consider the NFNS junction in the $x-y$ plane with the superconducting electrode covering the S region $x>L$. The ferromagnetic exchange energy $h$ is introduced in the F region $-L_f<x<0$ by the proximity effect. The two normal regions N$_1$ and N$_2$ are at $x<-L_f$ and $0<x<L$, respectively, as shown in Fig.\ \ref{Fig:1}. The electron and the hole excitations can be described by the Bogoliubov-de Gennes (BdG) equation \cite{PhysRevB.25.4515, de2018superconductivity}
\begin{equation}
\begin{pmatrix}
\mathcal{H}_{\sigma}-E_F+U(x) & \sigma\Delta_{SC}(x) \\
\sigma\Delta^*_{SC}(x) & E_F-U(x)-\mathcal{H}_{\bar{\sigma}}
\end{pmatrix}
\begin{pmatrix}
u_{\sigma} \\ v_{\bar{\sigma}}
\end{pmatrix}=\varepsilon\begin{pmatrix}
u_{\sigma} \\ v_{\bar{\sigma}}
\end{pmatrix},\label{Eq:3}
\end{equation}
where $u_{\sigma}$ and $v_{\bar{\sigma}}$ are the two-component electron spinor and the hole spinor respectively, $\varepsilon$ is the excitation energy measured from the Fermi energy $E_F$, $\sigma=\pm$ denotes the spin-up and the spin-down electrons respectively. The 2$\times$2 single particle Hamiltonian is $\mathcal{H}_{\sigma}=\mathcal{H}_0(\bm r)-\sigma h(x)$ with $h(x)=h\Theta(-x)\Theta(x+L_f)$. The $s$-wave superconducting pair potential $\Delta_{SC}(x)$ is zero in the normal and the ferromagnetic regions and is $\Delta_{SC}$ in the superconducting region, which can be induced by the proximity effect. The electrostatic potential $U(x)=-U_0\Theta(x-L)$ can be tuned by a gate voltage or by doping. We assumed $U_0+E_F\gg\Delta_{SC}$ so that the mean-field requirement in the superconducting region is fulfilled.

By solving the  BdG equation, we obtain the energy dispersions in the normal and the F region as
\begin{equation}
\begin{split}
\varepsilon_{e\sigma}=&\pm\sqrt{(k_x^2-\Delta_g)^2+k_y^2}-E_F-\sigma h(x),\\
\varepsilon_{h\bar{\sigma}}=&\pm\sqrt{(k_x^2-\Delta_g)^2+k_y^2}+E_F-\sigma h(x).\label{Eq:4}
\end{split}
\end{equation}
As shown in Fig.\ \ref{Fig:1}, the conduction band and the valence band meet at two Dirac points located at $\bm k=(\pm\sqrt{\Delta_g},0)$, which are labeled as $K$ and $K^{\prime}$ points. There are four propagating modes near the Dirac points. In the F region, the exchange energy $h$ is set to be larger than the inverted energy gap $\Delta_g$, and two of four propagating modes transform into two evanescent modes. In the F region, the eigenfunctions for the electron states in two Dirac cones are given as
\begin{equation}
\begin{split}
\Psi_{K\sigma}^{e\pm}=&\begin{pmatrix}
1 & \dfrac{k_{e\sigma\pm}^2-\Delta_g+i k_y}{\varepsilon+E_F+\sigma h} & 0 & 0
\end{pmatrix}^{\top}\exp(i k_{e\sigma\pm}x)/\sqrt{v_{\sigma\pm}},\\
\Psi_{K^{\prime}\sigma}^{e\pm}=&\begin{pmatrix}
1 & \dfrac{k_{e\sigma\mp}^2-\Delta_g+i k_y}{\varepsilon+E_F+\sigma h} & 0 & 0
\end{pmatrix}^{\top}\exp(-i k_{e\sigma\mp}x)/\sqrt{v_{\sigma\mp}},\label{Eq:5}
\end{split}
\end{equation}
where $k_{e\sigma\pm}=\left(\Delta_g\pm s_{e\sigma}\sqrt{(\varepsilon+E_F+\sigma h)^2-k_y^2}\right)^{1/2}$ is the longitudinal wave vector for the electron states in the Dirac cone with $s_{e\sigma}=\mathrm{sgn}(\varepsilon+E_F+\sigma h)$ and $v_{\sigma\pm}$ are given as
\begin{equation}
v_{\sigma\pm}=|k_{e\sigma\pm}(k_{e\sigma\pm}^2-\Delta_g)/(\varepsilon+E_F+\sigma h)|.\label{Eq:6}
\end{equation}
The eigenfunctions for the hole states in two Dirac cones are given as
\begin{equation}
\begin{split}
\Psi_{K^{\prime}\bar{\sigma}}^{h\pm}=&\begin{pmatrix}
0 & 0 & 1 & \dfrac{k_{h\bar{\sigma}\pm}^2-\Delta_g+i k_y}{E_F-\varepsilon-\sigma h}
\end{pmatrix}^{\top}\exp(i k_{h\bar{\sigma}\pm}x)/\sqrt{v_{h\bar{\sigma}\pm}},\\
\Psi_{K\bar{\sigma}}^{h\pm}=&\begin{pmatrix}
0 & 0 & 1 & \dfrac{k_{h\bar{\sigma}\mp}^2-\Delta_g+i k_y}{E_F-\varepsilon-\sigma h}
\end{pmatrix}^{\top}\exp(-i k_{h\bar{\sigma}\mp}x)/\sqrt{v_{h\bar{\sigma}\mp}},\label{Eq:7}
\end{split}
\end{equation}
where $k_{h\bar{\sigma}\pm}=\left(\Delta_g\pm s_{h\bar{\sigma}}\sqrt{(\varepsilon-E_F+\sigma h)^2-k_y^2}\right)^{1/2}$ is the longitudinal wave vector for the hole states in the Dirac cone with $s_{h\bar{\sigma}}=\mathrm{sgn}(\varepsilon-E_F+\sigma h)$ and $v_{h\bar{\sigma}\pm}$ are given as
\begin{equation}
v_{h\bar{\sigma}\pm}=|k_{h\bar{\sigma}\pm}(k_{h\bar{\sigma}\pm}^2-\Delta_g)/(\varepsilon-E_F+\sigma h)|.\label{Eq:8}
\end{equation}
By setting $h=0$, one can get the wave functions $\Psi_{K^{(\prime)}}^{e(h)\pm}$ and the longitudinal wave vector $k_{e(h)\pm}$ for the electron and the hole states in the N$_1$ and the N$_2$ regions.

In the superconducting region, the eigenfunctions are given by
\begin{equation}
\begin{split}
\Psi_{S1}=&\begin{pmatrix}
1 & 1 & e^{-i\beta} & e^{-i\beta}
\end{pmatrix}^{\top}\exp\left(i(k_0+i\kappa)x\right),\\
\Psi_{S2}=&\begin{pmatrix}
1 & -1 & e^{-i\beta} & -e^{-i\beta}
\end{pmatrix}^{\top}\exp\left(i(-\kappa+i k_0)x\right),\\
\Psi_{S3}=&\begin{pmatrix}
1 & 1 & e^{i\beta} & e^{i\beta}
\end{pmatrix}^{\top}\exp\left(i(-k_0+i\kappa)x\right),\\
\Psi_{S4}=&\begin{pmatrix}
1 & -1 & e^{i\beta} & -e^{i\beta}
\end{pmatrix}^{\top}\exp\left(i(\kappa+i k_0)x\right),\label{Eq:9}
\end{split}
\end{equation}
with $k_0=\sqrt{E_F+U_0}$ and $\kappa=\Delta_{SC}\sin\beta/(2k_0)$. The phase parameter is given by $\beta=\arccos(\varepsilon/\Delta_{SC})$ for $\varepsilon<\Delta_{SC}$ and $\beta=-i \mathrm{arccosh}(\varepsilon/\Delta_{SC})$ for $\varepsilon>\Delta_{SC}$.

In the junction, the eigenfunctions in the N$_1$ and the N$_2$ regions are the same. A wave incident on the F region is described in the basis of the eigenfunctions $\Psi_{K^{(\prime)}}^{e(h)\pm}$ by a vector of coefficients
\begin{equation}
a\equiv\begin{pmatrix}
a_{KL} & a_{K^{\prime}L} & a_{K^{\prime}R} & a_{KR} & a^h_{K^{\prime}L} & a^h_{KL} & a^h_{KR} & a^h_{K^{\prime}R}
\end{pmatrix}^{\top}.\label{Eq:10}
\end{equation}
The notation $a_{\tau\rho}$ and $a_{\tau\rho}^{(h)}$ denote the incident electron and the hole from the valley $\tau=K, K^{\prime}$ with the spin index being omitted. The index $\rho=L, R$ represents the N$_1$ and the N$_2$ regions respectively. The reflected and transmitted wave has a vector of coefficients 
\begin{equation}
b\equiv\begin{pmatrix}
b_{K^{\prime}L} & b_{KL} & b_{KR} & b_{K^{\prime}R} & b^h_{KL} & b^h_{K^{\prime}L} & b^h_{K^{\prime}R} & b^h_{KR}
\end{pmatrix}^{\top}.\label{Eq:11}
\end{equation}
The notation $b_{\tau\rho}$ and $b_{\tau\rho}^{(h)}$ denote the reflected and transmitted electron and hole, respectively.

The incident and the scattered waves at two interfaces of the F region can be related by $b=s_1a$. Since the F region does not couple electrons and holes, the matrix $s_1$ has the block-diagonal form
\begin{equation}
s_1(\varepsilon)=\begin{pmatrix}
s_e(\varepsilon) & 0 \\
0 & s_h(\varepsilon)
\end{pmatrix},\label{Eq:12}
\end{equation}
where $s_e$ is the 4$\times$4 unitary scattering matrix for electrons, while $s_h$ is for the hole part. $s_e$ and $s_h$ is given in a form as
\begin{equation}
s_e=\begin{pmatrix}
r & t^{\prime} \\
t & r^{\prime}
\end{pmatrix},s_h=\begin{pmatrix}
r_h & t_h^{\prime} \\
t_h & r_h^{\prime}
\end{pmatrix}. \label{Eq:13}
\end{equation}
The transmitted electron in the N$_2$ region is Andreev reflected as a hole at the NS interface. This process can be described by $a_R=s_2b_R$ with $a_R, b_R$ being expressed by
\begin{equation}
\begin{split}
a_R&=
\begin{pmatrix}
a_{K^{\prime}R} & a_{KR} & a^h_{KR} & a^h_{K^{\prime}R}
\end{pmatrix}^\top, \\
b_R&=
\begin{pmatrix}
b_{KR} & b_{K^{\prime}R} & b^h_{K^{\prime}R} & b^h_{KR}
\end{pmatrix}^\top. \label{Eq:14}
\end{split}
\end{equation}
The value of the matrix $s_1$ and $s_2$ can be obtained using the scattering matrix method \cite{PhysRevB.46.12841, PhysRevB.75.045426, RevModPhys.69.731}.

At subgap energy incidence, an electron incident from the left side of the junction is completely reflected. This process can be denoted by combining $s_1$ and $s_2$, leading to
\begin{equation}
\begin{split}
b_L&=\mathcal{R}a_L, \\
\mathcal{R}&=
\begin{pmatrix}
\mathcal{R}_{ee} & \mathcal{R}_{eh} \\
\mathcal{R}_{he} & \mathcal{R}_{hh}
\end{pmatrix}, \\
a_L&=
\begin{pmatrix}
a_{KL} & a_{K^{\prime}L} & a^h_{K^{\prime}L} & a^h_{KL}
\end{pmatrix}^\top, \\
b_L&=
\begin{pmatrix}
b_{K^{\prime}L} & b_{KL} & b^h_{KL} & b^h_{K^{\prime}L}
\end{pmatrix}^\top, \label{Eq:15}
\end{split}
\end{equation}
where $\mathcal{R}_{ee} (\mathcal{R}_{hh})$ is a $2\times2$ matrix represents the normal reflection for incident electron (hole) states, while $\mathcal{R}_{he} (\mathcal{R}_{eh})$ denotes the conversion from electron (hole) states to hole (electron) states. 

The differential conductance of the junction at zero temperature and subgap incident energy is given by $G/G_0=\sum_{\sigma}G_{\sigma}/G_0$ with
\begin{equation}
\begin{split}
G_{\sigma}=&\dfrac{e^2}{h}\int_{-k_{yc}}^{k_{yc}}\dfrac{d k_y}{2\pi/W}\mathrm{Tr}(1-\mathcal{R}_{ee}^{\dag}\mathcal{R}_{ee}+\mathcal{R}_{he}^{\dag}\mathcal{R}_{he})\\
=&\dfrac{e^2}{h}\int_{-k_{yc}}^{k_{yc}}\dfrac{d k_y}{2\pi/W}\mathrm{Tr}(2\mathcal{R}_{he}^{\dag}\mathcal{R}_{he}),\\
G_0=&\dfrac{4e^2}{h}\dfrac{k_{yc}W}{\pi}, k_{yc}=\varepsilon+E_F.\label{Eq:16}
\end{split}
\end{equation}
The factor 4 comes from 2 spin degeneracy and 2 forward propagating modes, and $W$ is the width of the junction.

\section{Discussion}

We first consider the NFN structure. As shown in Fig.\ \ref{Fig:1}, one can expect two valleys ($K$ and $K^{\prime}$) coexist when the condition $\Delta_g>(\varepsilon, E_F)$ is fulfilled. Under this condition, there exists four types of scattering processes for the electron incident from the $K$ valley, which are the intravalley  (intervalley) reflection and the intravalley (intervalley) transmission. The electron incidence is labeled by $a_K$ in Fig.\ \ref{Fig:1}, where the backscattering of $a_K\to b_K$ is the intravalley reflection and the one of $a_K\to b_{K^{\prime}}$ is the intervalley reflection.

 \begin{figure}[htbp]
\centering
\includegraphics[width=\linewidth]{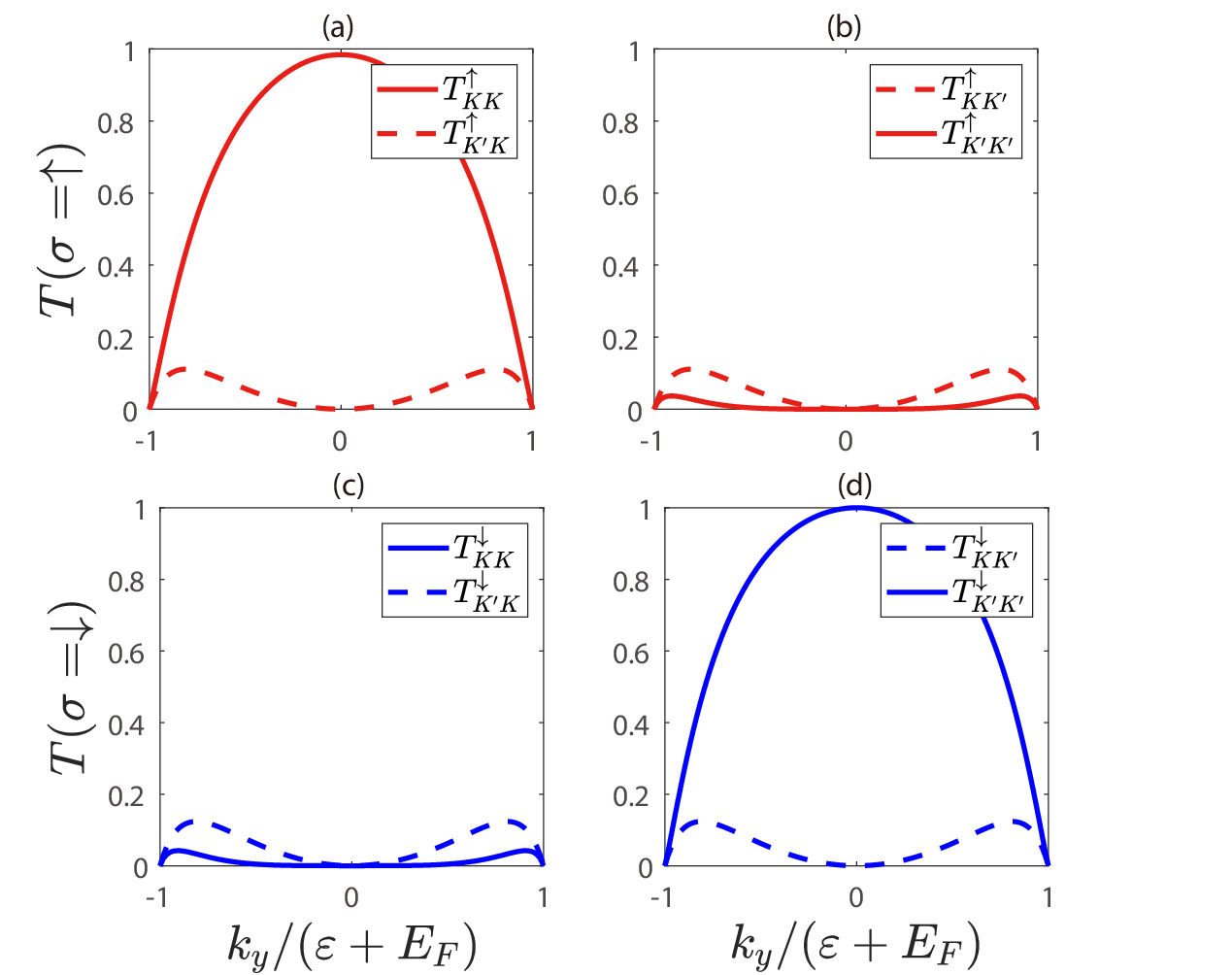}
\caption{\label{Fig:2}The transmission probability as functions of $k_y/(\varepsilon+E_F)$. The parameters are $\Delta_g=30$, $\varepsilon=0.98$, $L_f=2$, $E_F=0$, and $h=60$.
}
\end{figure}

The electron transport through the F region can be described by the scattering matrix $s_e$ in Eq.\ (\ref{Eq:13}). The submatrix $r$ and $t$ in $s_e$ are expressed by
 \begin{equation}
r=
\begin{pmatrix}
r^{\sigma}_{K^{\prime}K} & r^{\sigma}_{K^{\prime}K^{\prime}} \\
r^{\sigma}_{KK} & r^{\sigma}_{KK^{\prime}}
\end{pmatrix},
t=
\begin{pmatrix}
t^{\sigma}_{KK} & t^{\sigma}_{KK^{\prime}} \\
t^{\sigma}_{K^{\prime}K} & t^{\sigma}_{K^{\prime}K^{\prime}}
\end{pmatrix}, \label{Eq:17}
\end{equation}
where $r^{\sigma}_{\tau\tau^{\prime}}$ ($t^{\sigma}_{\tau\tau^{\prime}}$) denotes the reflection (transmission) amplitude from valley $\tau^{\prime}$ in N$_1$ region to valley $\tau$ in N$_2$ region with the spin index $\sigma=\uparrow, \downarrow$. We give the transmission probabilities $T^{\sigma}_{\tau\tau^{\prime}}=|t^{\sigma}_{\tau\tau^{\prime}}|^2$ as functions of $k_y$ in Fig.\ \ref{Fig:2}. For the spin-up electron incident from the $K$ valley, the intravalley transmission probability is nearly unity at normal incidence, and decreases monotonically down to 0 as $k_y$ increases, which is Klein-like. The transmission is intravalley-dominated, though the intervalley transmission is not strictly forbidden, with the probability being approximately $0.1$. For the incidence from the $K^{\prime}$ valley, the electrons are essentially reflected, and both the intervalley and the intravalley transmissions are negligible. For the spin-down electron incidence, the case is opposite. The electrons incident from the $K$ valley are fundamentally reflected, and the transmission from valley $K^{\prime}$ to $K^{\prime}$ is dominated. In short, for spin-up incidence, only the electron incident from the $K$ valley can transmit through the F region, while for spin-down incidence, only the electron from the $K^{\prime}$ valley can transmit throught the F region, i.e., the transport in NFN structure is spin-valley-polarized.

\begin{figure}[htbp]
\centering
\includegraphics[width=\linewidth]{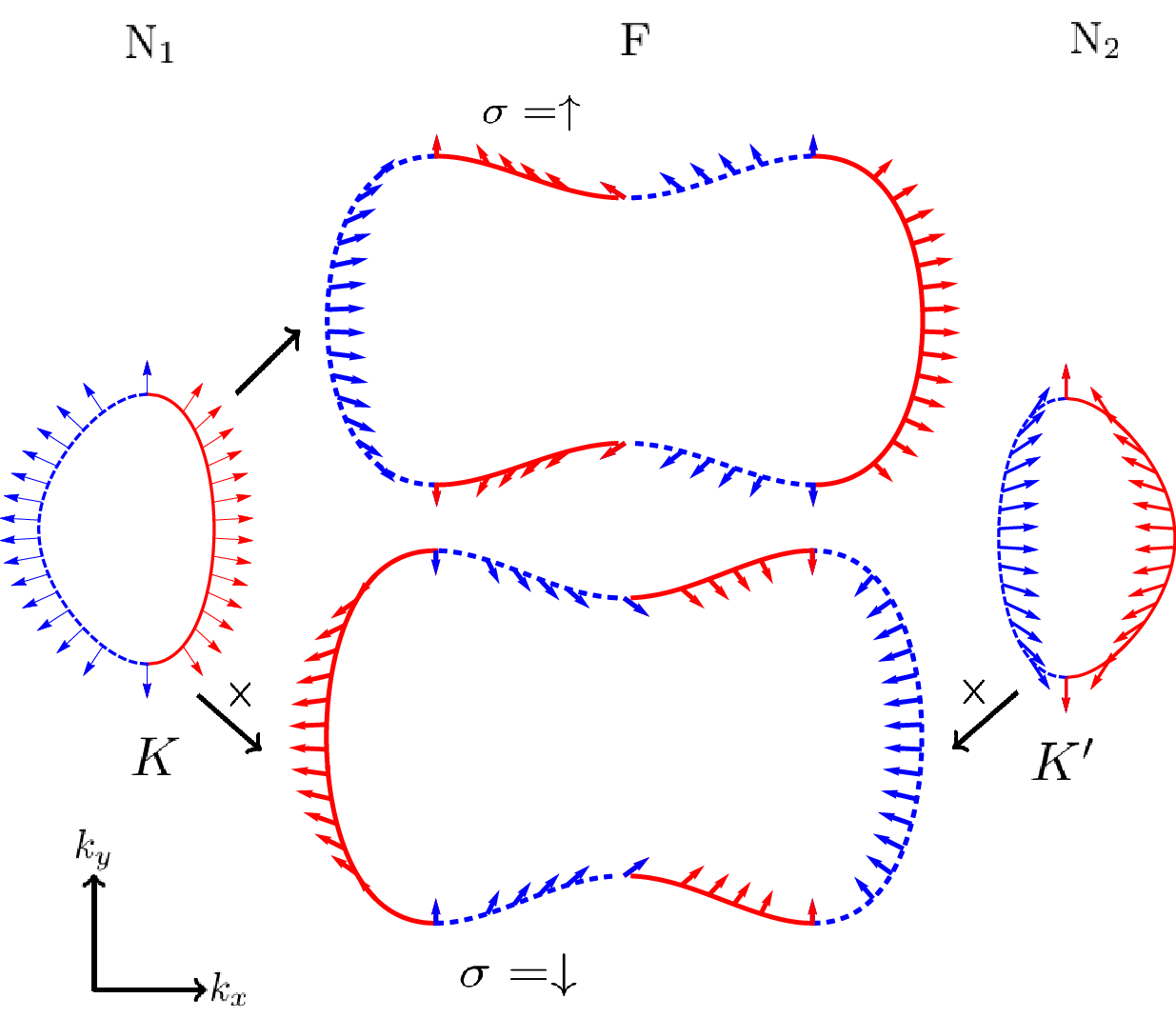}
\caption{\label{Fig:1(b)}The pseudospin texture of the particle. In the F region, the pseudospin of the spin-up subband is given in the upward panel, and that of the spin-down subband is given in the downward panel. The red solid and the blue dashed lines denote the forward and backward propagating states, respectively.
}
\end{figure}

The mechanism of the spin-valley polarization lies on the pseudospin texture, which is plotted in Fig.\ \ref{Fig:1(b)}. The pseudospin configuration in the normal region is split into two parts corresponding to two valleys. In the F region, $h$ is larger than $\Delta_g$, and the two parts merge into a single one owing to the lifting of the Fermi level. The pseudospin texture is defined as $\bm P=(P_x, P_y)=\Psi^{\dagger}\bm s\Psi$ with $\bm s$ being the Pauli matrix in the pseudospin space. The pseudospin texture of the propagating states is spin-degenerate in the normal region but spin-resolved in the F region. As shown in Fig.\ \ref{Fig:1(b)}, for the spin-up electron incident from the $K$ valley in the N$_1$ region, the pseudospin of the forward propagating states matches that of the forward states in the F region, resulting in the transmission of the spin-up electrons. On the contrary, for the spin-down electron, the pseudospin of the forward propagating states mismatches that of the  forward states in the F region, hence the transmission of the spin-down electron is suppressed. The case in the $K^\prime$ valley is just opposite. Therefore, there are only spin-up current in the $K$ valley and spin-down current in the $K^\prime$ valley resulting in the spin-valley polarization. When the direction of the magnetization is reversed, the direction of the pseudospin in the F region is reversed as well, which leads to the switching effect of the spin polarization in two valleys.

\begin{figure}[htbp]
\centering
\includegraphics[width=\linewidth]{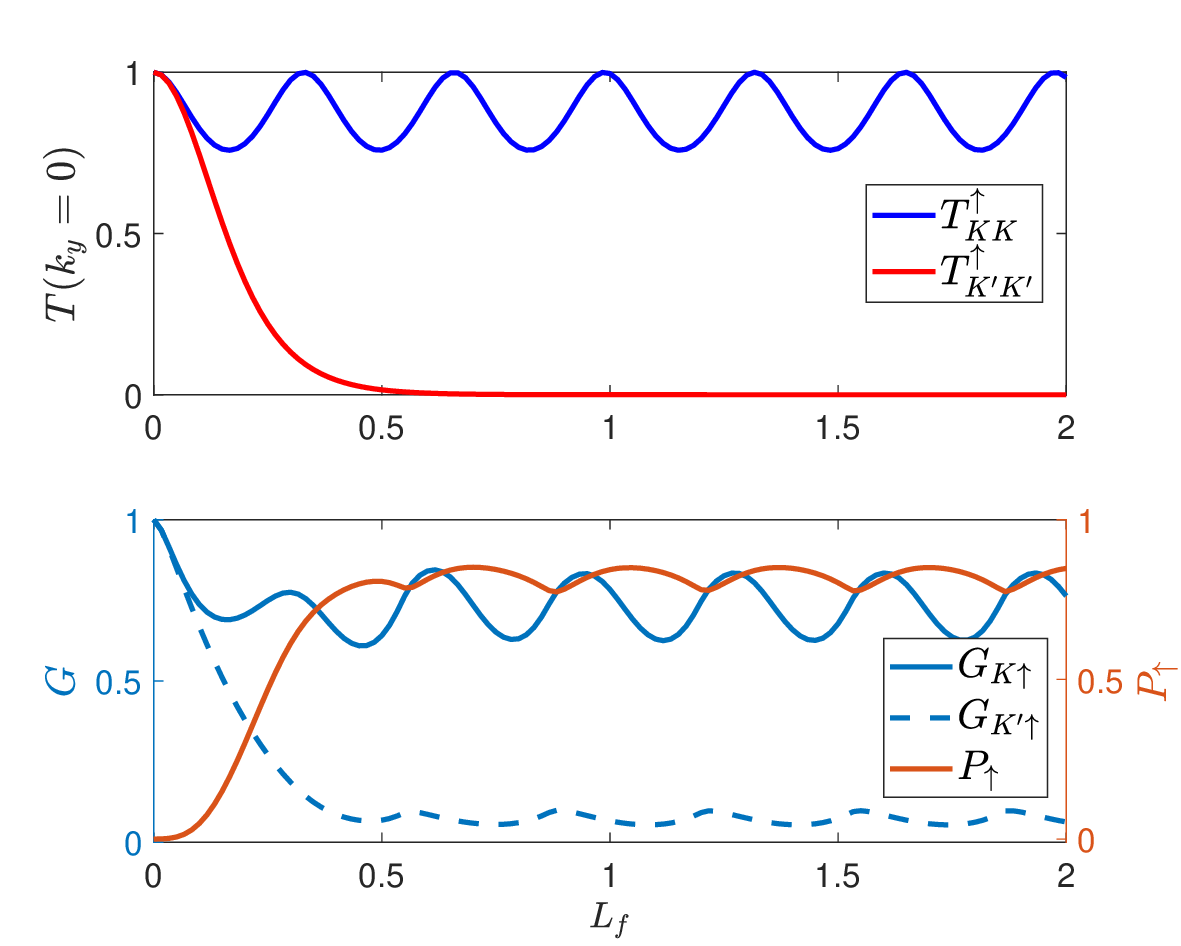}
\caption{\label{Fig:4}(a)The probability $T(k_y=0)$ for intravalley transmission in two valleys. (b)The valley resolved conductance and the valley polarization for spin-up incidence as functions of $L_f$. The parameters are $\Delta_g=30$, $\varepsilon=0.98$, $E_F=0$, and $h=60$.
}
\end{figure}

We define the valley polarization $P_{\uparrow}$ for spin-up electrons
\begin{equation}
\begin{split}
P_\uparrow&=\dfrac{G_{K\uparrow}-G_{K^{\prime}\uparrow}}{G_{K\uparrow}+G_{K^{\prime}\uparrow}}, \\
G_{K\uparrow}&=\int_{-k_{yc}}^{k_{yc}}dk_y\sum_{\tau=K, K^{\prime}}|t^{\uparrow}_{K\tau}|^2, \\
G_{K^{\prime}\uparrow}&=\int_{-k_{yc}}^{k_{yc}}dk_y\sum_{\tau=K, K^{\prime}}|t^{\uparrow}_{K^{\prime}\tau}|^2, \label{Eq:18}
\end{split}
\end{equation}
where $G_{K\uparrow}$ and $G_{K^{\prime}\uparrow}$ are the valley resolved conductances for spin-up electrons. The transmission probability at normal incidence, which can be seen as the single channel conductance at $k_y=0$, is shown in Fig.\ \ref{Fig:4}(a). The conductance $G_{K\uparrow}$, $G_{K^{\prime}\uparrow}$ and the valley polarization $P_\uparrow$ as functions of $L_f$ for spin-up electron incidence is shown in Fig.\ \ref{Fig:4}(b). It is shown that the intravalley transmission probability $T^{\uparrow}_{KK}$ for normal incidence oscillates with $L_f$ due to the interference effect, with its peak value being 1. For the transmission from valley $K^{\prime}$ to $K^{\prime}$, $T^{\uparrow}_{K^{\prime}K^{\prime}}$ decreases to zero as $L_f$ increases. To better understand the behavior of the curves, we give the analytical form of the probability by solving Eqs.\ (\ref{Eq:5})-(\ref{Eq:13}), which is given as
\begin{equation}
\begin{split}
T&_{KK}^{\uparrow}(k_y=0)=\dfrac{2}{\gamma_++1-(\gamma_+-1)\cos(2q_+L_f)}, \\
T&_{K^{\prime}K^{\prime}}^{\uparrow}(k_y=0)=\dfrac{1}{1+\gamma_- \mathrm{sinh}^2(q_-L_f)},\\
\gamma&_{\pm}=(k_{\pm}^2+q_{\pm}^2)^2/(4k_{\pm}^2q_{\pm}^2),
\label{Eq:19} 
\end{split}
\end{equation}
where $q_{\pm}=\sqrt{|\varepsilon+E_F+h|\pm\Delta_g}$ and $k_{\pm}=\sqrt{\Delta_g\pm|\varepsilon+E_F|}$.  As $L_f$ increases, the term $\mathrm{sinh}^2(q_-L_f)$ tends  to infinity, and $T_{K^{\prime}K^{\prime}}^{\uparrow}(k_y=0)$ decays to zero. When we integrate over the transverse momentum $k_y$, the curve of $G_{K\uparrow}$ is similar to that of $T_{KK}^{\uparrow}(k_y=0)$. $G_{K^{\prime}\uparrow}$ decays with the increasing $L_f$ but remains finite even for large $L_f$. The finite $G_{K^{\prime}\uparrow}$ is due to the incomplete pseudospin forbidden transition at oblique incidence. As shown in Fig.\ \ref{Fig:4}(b), $P_\uparrow$ first increases sharply from 0 to approximately 0.85 and finally exhibits a oscillatory behavior as $L_f$ increases.

For the spin-down incidence, the case is the opposite. Consequently, the valley polarization in the spin-up subband is opposite to that in the spin-down subband, exhibiting the so-called spin-valley-polarization.

\begin{figure}[htbp]
\centering
\includegraphics[width=\linewidth]{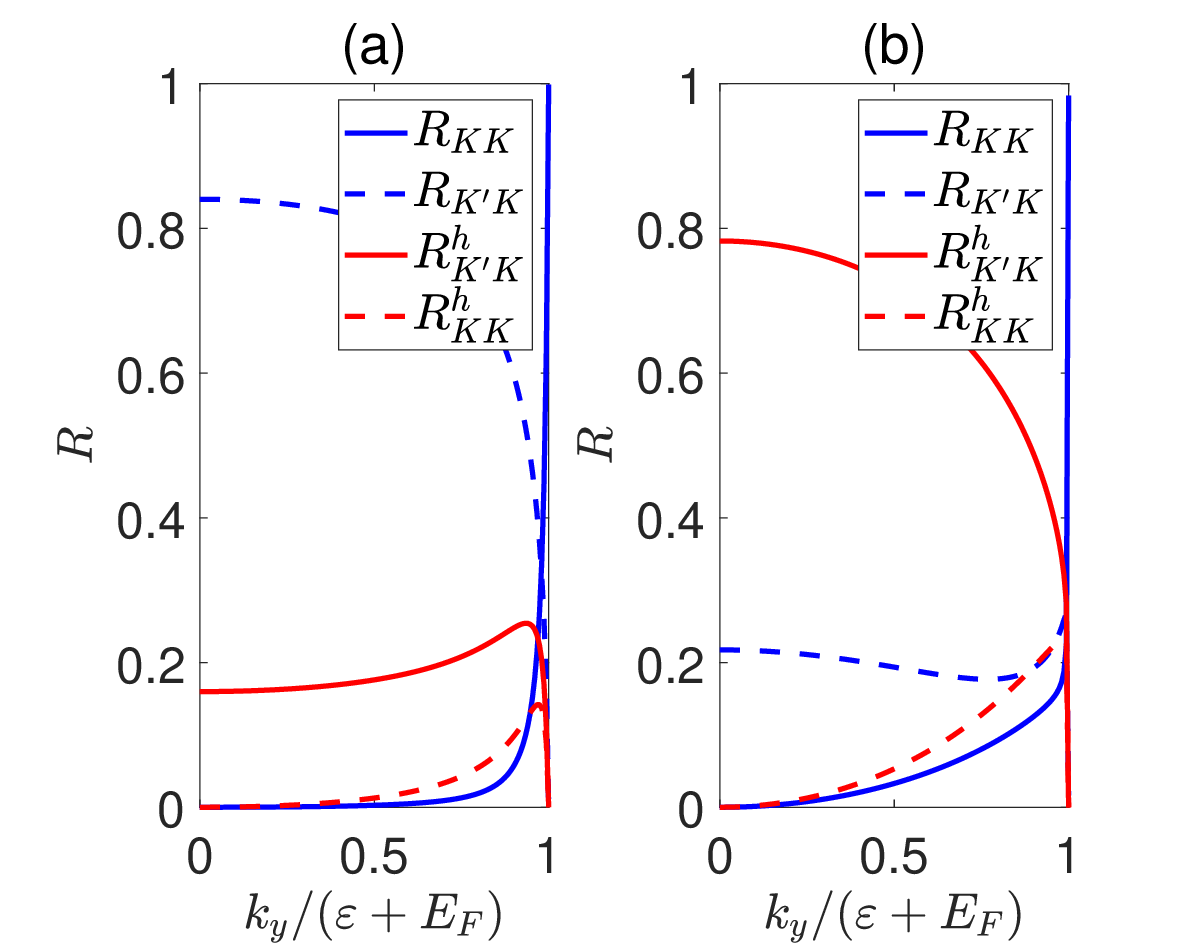}
\caption{\label{Fig:5}Reflection coefficients as a function of $k_y/(\varepsilon+E_F)$. The Fermi energy $E_F$ is set to be 0, the electrostatic potential $U_0$ is 900, the superconducting pair potential $\Delta_{SC}$ is 1, the inverted energy gap $\Delta_g$ is 30. $\varepsilon$ is set to be 0.5 in (a) and 0.98 in (b).
}
\end{figure}

Now we consider the NS interface in the junction. There are four types of scattering processes at the NS interface, which are the intravalley (intervalley) normal reflection $R_{KK}$ ($R_{K^{\prime}K}$) and the intravalley (intervalley) AR $R^h_{KK}$ ($R^h_{K^{\prime}K}$). The reflection probabilities  in a pure NS junction are given in Fig.\ \ref{Fig:5} with the incident energy $\varepsilon$ being 0.5$\Delta_{SC}$ and 0.98$\Delta_{SC}$, respectively. When $\varepsilon$ is 0.5, the intravalley AR probability is negligibly small in the most range of incident angles due to the momentum mismatch and the intervalley AR is dominant. The ratio of two types of AR satisfies $R^h_{KK}/R^h_{K^{\prime}K}\propto\cot^2(\varphi/2)$ with $\varphi=\arcsin k_y/(E_F-\varepsilon)$. As $k_y/(\varepsilon+E_F)$ goes from 0 to 1, $\varphi$ changes from $\pi$ to $3\pi/2$, so that the probability of intravalley AR is always smaller than that of the intervalley AR, especially when $k_y$ is small. The intervalley AR probability at normal incidence is evaluated as
\begin{equation}
R^h_{K^{\prime}K}(k_y=0)\simeq\dfrac{k_{e+}^2}{k_0^2\sin^2\beta+(k_{e+}+k_{h-})^2\cos^2\beta}. \label{Eq:20}
\end{equation}
Note that $k_0^2\gg(k_{e+}+k_{h-})^2$, and the probability increases monotonically as $\varepsilon$ approaches $\Delta_{SC}$. When $\varepsilon\to\Delta_{SC}$, $\beta$ tends to be 0, and the probability becomes $R^h_{K^{\prime}K}(k_y=0)\to k_{e+}^2/(k_{e+}+k_{h-})^2<1$. Different from electron scattering, the AR for normal incidence cannot reach unity.

\begin{figure}[htbp]
\centering
\includegraphics[width=\linewidth]{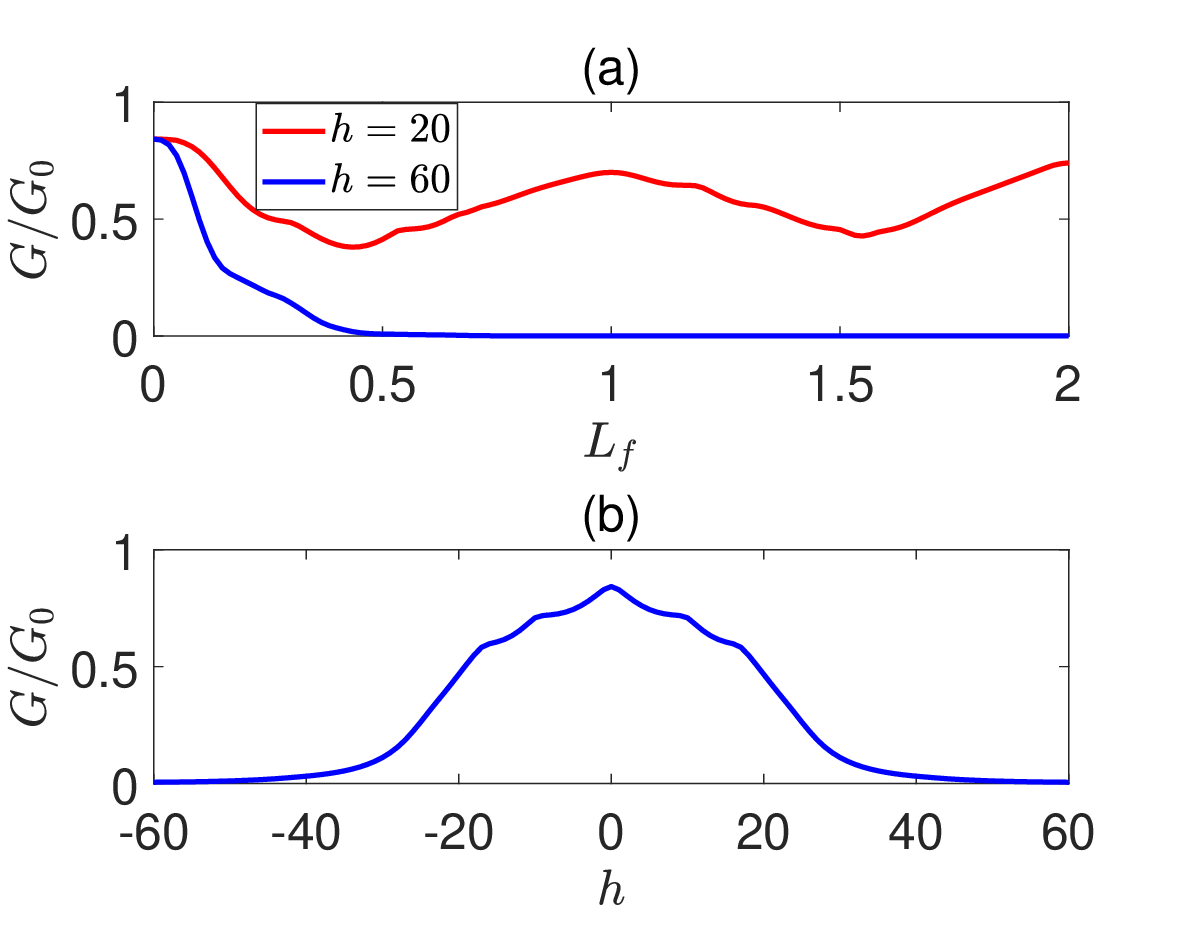}
\caption{\label{Fig:6}The normalized differential conductance $G/G_0$ as functions of $L_f$ and $h$. The parameters are $\Delta_g=30$, $\varepsilon=0.98$, $E_F=0$, and $\Delta_{SC}=1$ respectively. In (a), $h=20$ for the red line and $h=60$ for the blue line. In (b), $L_f=0.6$.
}
\end{figure}

Finally we take the whole NFNS junction into account. We calculate the differential conductance $G/G_0$ at zero temperature numerically as functions of $L_f$ and $h$ in Fig.\ \ref{Fig:6}(a) and Fig.\ \ref{Fig:6}(b) respectively. The length of the N$_2$ region is set to be much smaller compared to the length of the F region as well as the superconducting coherence length, so that it has little effect on the conductance. In Fig.\ \ref{Fig:6}(a), when $h<\Delta_g$, the conductance oscillates with the increase of $L_f$ without decaying, which stems from the failure of the valley-filtering effect. When $h>\Delta_g$, the conductance decreases sharply as $L_f$ increases and is strongly suppressed. One finds that the behavior of the conductance $G/G_0$ coincides with that of the valley polarization $P_\uparrow$ as a function of $L_f$. In Fig.\ \ref{Fig:6}(b), the conductance decreases monotonically in a oscillatory way with the increase of $h$, showing that the spin-valley polarization has a positive correlation with the value of $h$.

The nearly zero differential conductance can be explained as follows. A spin-up electron incident from the $K$ valley in the N$_1$ region is first transmitted through the F region, with the electron from the other valley being filtered. At the NS interface of the junction, the spin-up electron in the $K$ valley is Andreev reflected as a spin-down hole in the $K^{\prime}$ valley. In the F region, the pseudospin of the backward propagating states in the spin-down subband mismatches that of the $K^{\prime}$ valley, which leads to the inhibition of the transmission of the spin-down hole in the $K^{\prime}$ valley, as shown in Fig.\ \ref{Fig:1(b)}. Similarly, when we consider an electron incident from the $K^{\prime}$ valley, only the spin-down electron can transmit through the F region and is then Andreev reflected as a spin-up hole in the $K$ valley. The pseudospin of the backward propagating states in the spin-up subband mismatches that of the $K$ valley, hence the transport of the spin-up hole is suppressed.

\begin{figure}[htbp]
\centering
\includegraphics[width=\linewidth]{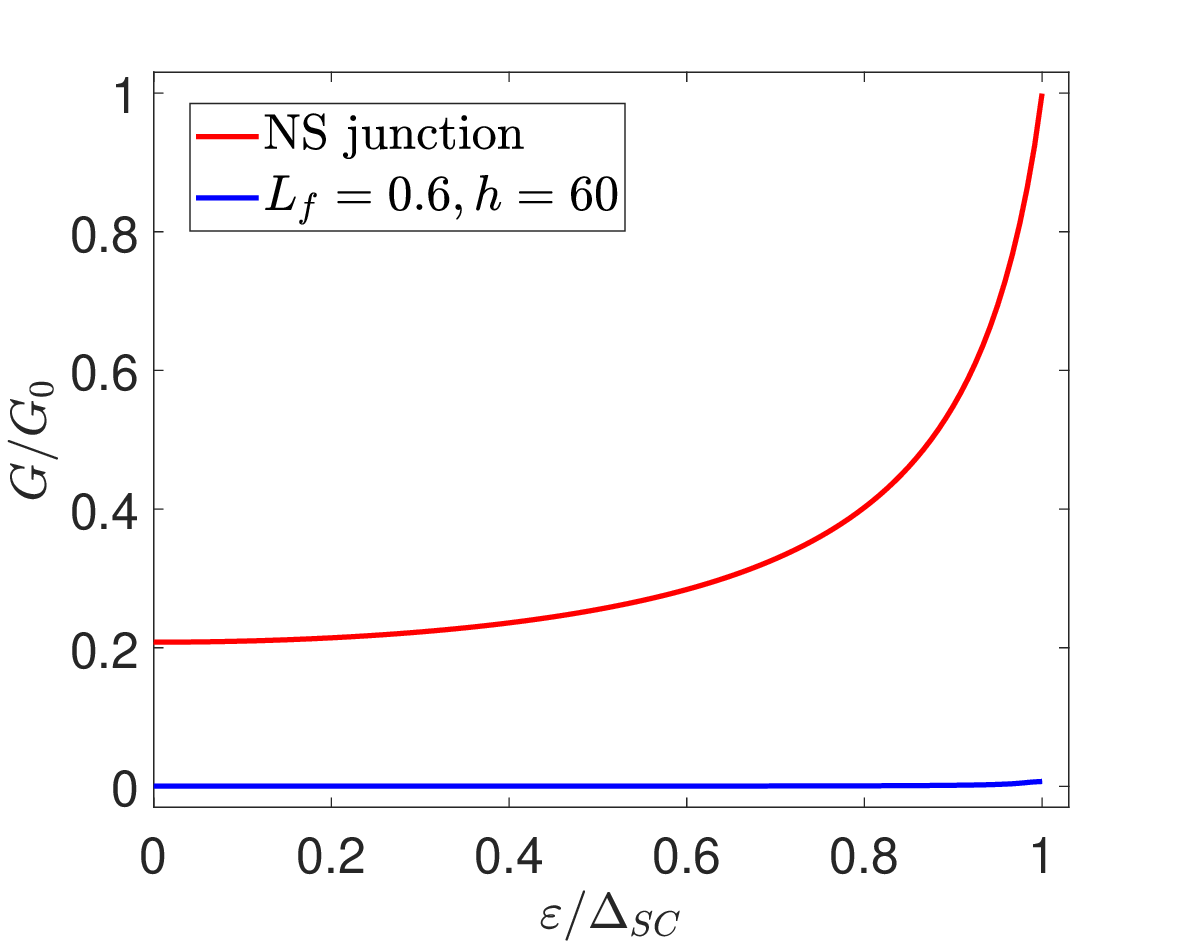}
\caption{\label{Fig:7}The normalized differential conductance $G/G_0$ as a function of the incident energy $\varepsilon/\Delta_{SC}$. The parameters are $\Delta_g=30$, $E_F=0$, and $\Delta_{SC}=1$. The red and blue curves represent $L_f=0, h=0$ (a pristine NS interface) and $L_f=0.6, h=60$, respectively.
}
\end{figure}
The differential conductance is plotted as a function of the incident energy in Fig.\ \ref{Fig:7}. In a pristine NS junction, the conductance increases monotonically from about 0.2 to 1. In the NFNS junction, the conductance is approximately zero. Though the total conductance of two valleys is not spin polarized, the Andreev conductance still vanishes as a clear signal for the spin-valley polarization.

\section{Conclusion}

In this work, we investigated the electron tunneling and the AR in the NFNS junction based on semi-Dirac materials with inverted energy gap. Due to the ferromagnetic exchange energy and the pseudo-spin texture in the semi-Dirac material, the electron transport in two valleys are opposite spin polarized. Such a spin-valley polarization can be indicated by the vanishing of the AR in the NFNS structure.

\begin{acknowledgments}

This work is supported by the National Key R\&D Program of China (Grant No. 2022YFA1403601).
\end{acknowledgments}

\end{document}